\documentclass[aps,prl,twocolumn,nobibnotes,superscriptaddress,notitlepage]{revtex4-2}

\usepackage[hyperindex,breaklinks,colorlinks,urlcolor=JungleGreen,citecolor=JungleGreen,linkcolor=JungleGreen]{hyperref}

\usepackage[utf8]{inputenc}
\usepackage[usenames,dvipsnames]{xcolor}
\usepackage{float}
\usepackage{dcolumn} 
\usepackage{bm}
\usepackage{graphicx}
\usepackage{mathrsfs}
\usepackage{overpic}
\usepackage{wrapfig}
\usepackage{natbib}
\usepackage{braket}
\usepackage{color}
\usepackage{verbatim}
\usepackage{amssymb,amsmath,mathtools}
\usepackage{upgreek}
\usepackage{bbold}
\usepackage{cancel}
\usepackage{textgreek}
\usepackage{dsfont}
\usepackage[normalem]{ulem}
\usepackage{hyperref} 

\definecolor{darkblue}{rgb}{0.0 0.0 0.78}
\definecolor{darkred}{rgb}{0.5 0.0 0.0}

\newcommand{\UMDphy}{Department of Physics, University of Maryland, College Park, Maryland 20742, USA}
\newcommand{\QTC}{Quantum Technology Center, University of Maryland, College Park, Maryland 20742, USA}
\newcommand{\UMDEECS}{Department of Electrical and Computer Engineering, University of Maryland, College Park, Maryland 20742, USA}

\newcommand{\ARL}{DEVCOM Army Research Laboratory, Adelphi, Maryland 20783, USA}

\begin{document}

\title{Beyond Average Hamiltonian Theory for Quantum Sensing}
\date{\today}

\author{Jner Tzern Oon}
\affiliation{\UMDphy}
\affiliation{\QTC}

\author{Sebastian C. Carrasco}
\affiliation{\ARL}

\author{Connor A. Hart}
\affiliation{\QTC}

\author{George Witt}
\affiliation{\UMDphy}

\author{Vladimir S. Malinovsky}
\affiliation{\ARL}

\author{Ronald Walsworth}
\affiliation{\UMDphy}
\affiliation{\QTC}
\affiliation{\UMDEECS}

\begin{abstract}

The application of average Hamiltonian theory (AHT) to magnetic resonance and quantum sensing informs pulse sequence design, for example, by providing efficient approximations of spin dynamics while retaining important physical characteristics of system evolution.
However, AHT predictions break down in many common experimental conditions, including for sensing with solid-state spins. 
Here we establish that certain symmetries, such as rapid echos, allow AHT to remain accurate well beyond the perturbative limit. 
An exact method is presented to determine the sensor response to a target signal, which stays valid beyond the regime of AHT convergence.
This beyond AHT approach enables new opportunities in quantum control techniques that leverage complementary analytical and numerical methods, with applications in a variety of quantum sensing platforms, Hamiltonian engineering, and probes of quantum many-body phenomena.

\end{abstract}
\date[]{}
\maketitle

\section{Introduction}
The Magnus expansion offers a general solution to a set of non-autonomous linear differential equations involving linear operators~\cite{Magnus1954}. As a result, it is naturally suited for studying a wide range of phenomena within physics,
finding frequent use in classical mechanics~\cite{Oteo1991}, quantum field theory~\cite{Connes2000}, atomic physics~\cite{Robinson1963, Eichler1977}, and elsewhere~\cite{Blanes2009}.
Considering its ubiquity, studies of the conditions under which the Magnus expansion is valid has widespread implications~\cite{Casas2007}.

Notably, the Magnus expansion forms the mathematical foundation of average Hamiltonian theory (AHT)~\cite{Brinkmann2016}, which was initially developed for studies of nuclear magnetic resonance (NMR)~\cite{Evans1968, Haeberlen1968}. 
By expressing the solution of the time-dependent Schr\"{o}dinger
equation in an exponential form, AHT  
approximates time evolution via an effective time-independent Hamiltonian obtained using the Magnus expansion~\cite{Magnus1954}.
Besides providing accurate numerical predictions in the perturbative limit, AHT retains hermiticity at any truncation order, while often preserving qualitative physical properties of system dynamics~\cite{Edn1999}. Over several decades, AHT has seen prolific use in the development of pulse sequences for decoupling of spin-spin interactions; and is a powerful tool for both analytical and numerical predictions of closed quantum dynamics~\cite{Goldman1992, Mananga2011, Candoli2023}. 
Experimentally, such pulse sequences typically involve the time-dependent application of resonant or near-resonant radio frequency, microwave, and/or optical fields suitable for manipulation of the internal (e.g., spin) states of the system of interest.

Recently, the application of AHT has been extended to the field of quantum sensing, where its use ranges from semi-quantitative analysis of pulse sequences for dipolar decoupling~\cite{Sahin2022, Balasubramanian2019} to forming the mathematical foundation of 
novel sensing sequences for strongly interacting spin systems~\cite{Choi2017, Choi2020, OKeeffe2019}.  Notably, research into the latter has contributed to experimental demonstrations of improved coherence times and magnetic field sensitivity for dense ensembles of nitrogen-vacancy (NV) centers in diamond~\cite{Zhou2020, Arunkumar2023}.

Here, we show that the conditions that guarantee Magnus expansion convergence conflicts with common operating conditions for magnetometry with solid-state spins (including NV-diamond), leading to a breakdown in AHT.
Beyond the regime of guaranteed convergence, we demonstrate that specific symmetries allow pulse sequences to retain Magnus expansion accuracy.
Moreover, we present an exact method to evaluate the sensor response to a target classical field, which bypasses the limitations of AHT. We demonstrate the utility of this exact method on widely used pulse sequences for both broadband (DC) and narrowband (AC) magnetometry.
This work opens up possibilities for pulse sequence engineering beyond the AHT regime; identification of further symmetries that preserve AHT convergence; and development of numerical techniques that accurately characterize sensor performance.

\section{Average Hamiltonian Theory}
To describe a general pulse sequence applied to a quantum system, we consider a series of $n$ control pulses with unitary operators $P_1, \cdots, P_n$. Each pulse is immediately followed by free evolution of the quantum system for duration $\tau_1, \dots, \tau_{n}$, respectively.  
With each applied pulse, an initial Hamiltonian $H_0$ is rotated to be
\begin{equation}
\label{eq:Htoggling}
H_i = (P_i \dots P_1)^\dagger H_0 (P_i \dots P_1),
\end{equation}
such that state evolution during the $i$-th free evolution frame is given by the piecewise constant Hamiltonian $H_i$. For a total sequence duration $t = \tau_1 + \tau_2 \dots 
 \tau_n$, 
AHT approximates the unitary propagator as \footnote{Multiplicative factors of the reduced Planck constant $\hbar$ are dropped in the exponent for simplicity. All Hamiltonian operators have non-angular frequency units in this work.} 
\begin{equation}
\label{eq:uni_aht}
\mathcal{U}^{(m)}(t) = \exp\left(- i \mathcal{H}^{(m)} t \right),
\end{equation}
where $\mathcal{H}^{(m)}$ denotes a time-independent effective Hamiltonian described by the Magnus expansion,
\begin{equation}
\label{eq:magnusseries}
\mathcal{H}^{(m)} = \bar H^{(1)} +  \bar H^{(2)}  \dots + \bar H^{(m)}.
\end{equation}
For reference, the first two terms in the expansion are provided in Eq.~(\ref{eq:Hterms}). The first order contribution $\bar H^{(1)}$ describes the mean of the interaction frame Hamiltonians $H_i$, weighted by the duration of each corresponding free evolution interval $\tau_i$. Higher order terms consist of summations over increasing numbers of nested commutators.
\begin{equation}
\begin{aligned}
\label{eq:Hterms}
\bar H^{(1)} &= \frac{\sum_i^n H_i \tau_i}{\sum_i^n \tau_i} \\
\bar H^{(2)} &= \frac{1}{2i\sum_{i=1}^{n} \tau_j} \sum_{j=1}^{n} \sum_{i=1}^{j-1} \big[H_j, H_i \big] \tau_j \tau_i
\end{aligned}
\end{equation}


\section{AHT Breakdown}

\begin{figure}[t]
    \centering
\includegraphics[width=\linewidth]{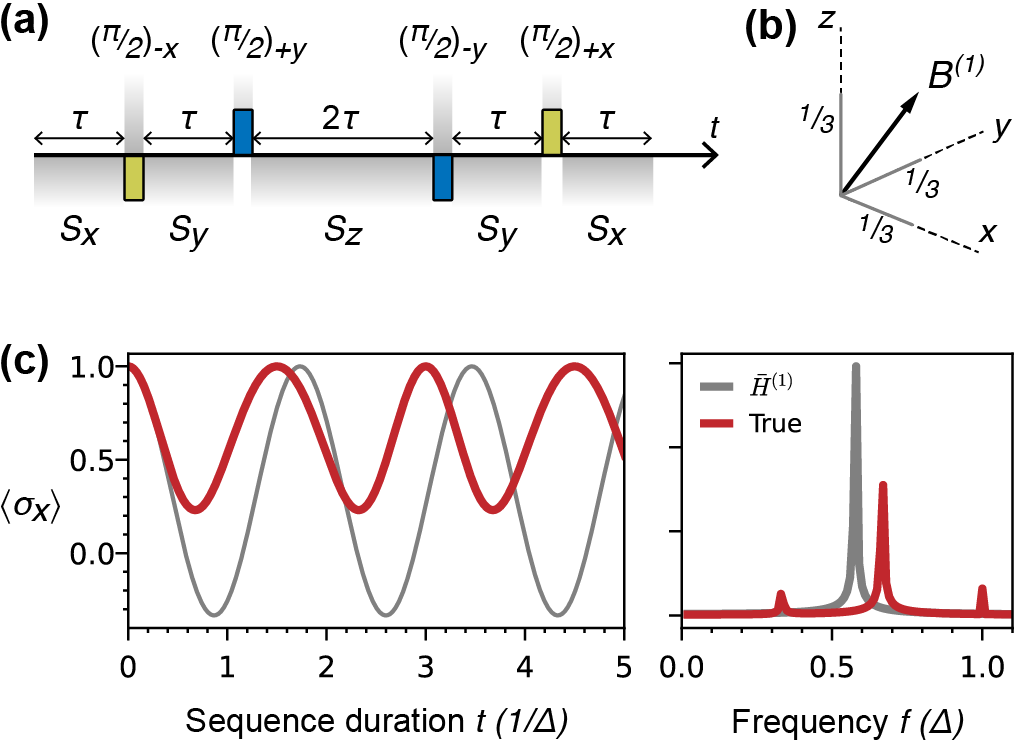}
    \caption{Fig 1
(a) WAHUHA pulse sequence and toggling frame representations. (b) Effective field corresponding to leading order average Hamiltonian.  (c) WAHUHA time series $\braket{\sigma_x(t)}$ and corresponding power spectra, obtained using the first order average Hamiltonian (gray) vs ground truth unitary evolution (red).  
}
    \label{fig:1}
\end{figure}

To characterize the effect of a specific pulse sequence on sensor performance,
we define the sensor coupling factor $\gamma=\partial\Delta/\partial V$, which represents a first order shift in the sensor energy level difference $\Delta$ with respect to a target (classical) signal field $V$ to be measured~\cite{Degen2017}. Sensitivity to changes in $V$ scales linearly with respect to $\gamma$, roughly following
$$\eta \propto \frac{1}{\gamma \sqrt{T_d}},$$
where $T_d$ refers to the characteristic timescale over which a measurement observable decays (for example, due to dephasing or decoherence). 

For accurate assessments of sensitivity, reliable estimates of $\gamma$ are crucial. 
For example, when using electronic spins for magnetic field measurements, $\gamma$ represents an ``effective" gyromagnetic ratio, and can be defined with respect to the electron gyromagnetic ratio $\gamma_e \approx - 2.8 \cdot 10^{10} \text{ Hz/T}$. 
We therefore introduce a dimensionless coupling factor $\alpha = \gamma/\gamma_e$, which represents a suppression of the bare electron response that depends on the sensing protocol employed.
In other words, $\alpha$ is fixed for a given sensing protocol, even if the sensor platform (e.g., electronic spins) is changed.

For a given sensing protocol, previous approaches utilized AHT to estimate $\alpha$~\cite{OKeeffe2019, Zhou2020}. To evaluate the sensor response due to a target Hamiltonian $H_0$, the first order average Hamiltonian $\mathcal{H}^{(1)}$ is projected onto the relevant basis operators, e.g., the spin operators $S_x, S_y, S_z$. The components of $\alpha$ (normalized by $H_0$) are calculated using
\begin{equation}
\label{eq:projcomponent}
a_{x,y,z} = \frac{\text{Tr}(\mathcal{H}^\dag S_{x,y,z})}{\text{Tr}(H_0^\dag H_0)},
\end{equation}
such that $\alpha = \sqrt{a_x^2 + a_y^2 + a_z^2}$.

To provide intuition for $\alpha$, we consider the highly influential WAHUHA pulse sequence, which was originally conceived by Waugh et al.~\cite{Waugh1968} to decouple homonuclear magnetic dipolar interactions in solid-state NMR. 
As shown in Fig \ref{fig:1}(a), the WAHUHA protocol consists of four $\pi/2$-rotation pulses applied along $-S_x, S_y, -S_y, S_x$, where $S_{x,y,z}$ denote spin operators along the $x,y,z$ directions. The pulses are accompanied by intervals of free evolution, expressed in units of time $\tau$. By spending equal time along all three orthogonal magnetic field directions, WAHUHA generates a vanishing first order average Hamiltonian $\bar H^{(1)} = 0$ for spins 
coupled by dipolar interactions. 


Applying WAHUHA to a target magnetic field $\vec{B}_0 = B_0\hat{z}$ such that $H_0 = - \gamma_e B_0 S_z$, Eq.~(\ref{eq:Hterms}) reveals the first order average Hamiltonian, 
\begin{equation}
\label{eq:WAHUHA_H1}
\mathcal{H}^{(1)} = -\gamma_e  \frac{B_0}{3}(S_x+S_y+S_z).
\end{equation}
Here, $\mathcal{H}^{(1)}$ is described by an effective field $\vec{B}^{(1)} = (B_0/3, B_0/3, B_0/3)$, as illustrated in Fig. \ref{fig:1}(b). 
Setting $\mathcal{H} = \mathcal{H}^{(1)}$ in Eq.~(\ref{eq:projcomponent}) results in $\alpha = |\vec{B}^{(1)}|/|\vec{B}_0| = 1/\sqrt{3}$. 
Indeed, $\alpha$ denotes the fractional length of the effective field associated with $\mathcal{H}$, relative to the corresponding field of the target Hamiltonian $H_0$ \footnote{Alternatively, Ref.~\cite{Zhou2024} calculates $\alpha$ by considering the largest difference between the eigenvalues of $\bar H^{(1)}$, which gives the same result for a two-level system.}. 

The leading order result in Eq.~(\ref{eq:WAHUHA_H1}) predicts state precession around the effective field $\vec{B}^{(1)}$ with a single-tone frequency $\Delta/\sqrt{3}$, where $H_0 = \Delta S_z$.
In Fig 1(c), we show the resulting spin dynamics due to $\mathcal{U}^{(1)}$
and compare to the time evolution using the exact unitary,
\begin{equation}
\label{eq:U0}
 U_0 =  \left(e^{- i H_0 \tau_n} P_n \right)\cdots \left(e^{- i H_0 \tau_2} P_2 \right) \left(e^{- i H_0 \tau_1} P_1\right).
\end{equation}
Here, we assume instantaneous pulses applied to an initial state prepared along $S_x$. 
The time-dependent state projection onto $S_x$ is plotted as a function of time $t$, evolved separately using unitaries $\mathcal{U}^{(1)}$ (gray curve) and $U_0$ (red curve). 
Immediately, we observe significant differences.
In contrast to the single frequency oscillation $\Delta/\sqrt{3}$ predicted by $\bar H^{(1)}$, the true time series exhibits beatings with multiple tones, with none at the frequency $\Delta/\sqrt{3}$. This behavior is also clearly seen in the accompanying power spectra in Fig. \ref{fig:1}(c).
The amplitude of the WAHUHA fringes also greatly differ from AHT predictions.

\section{Magnus Expansion Convergence}

\begin{figure}[t]
    \centering
\includegraphics[width=\linewidth]{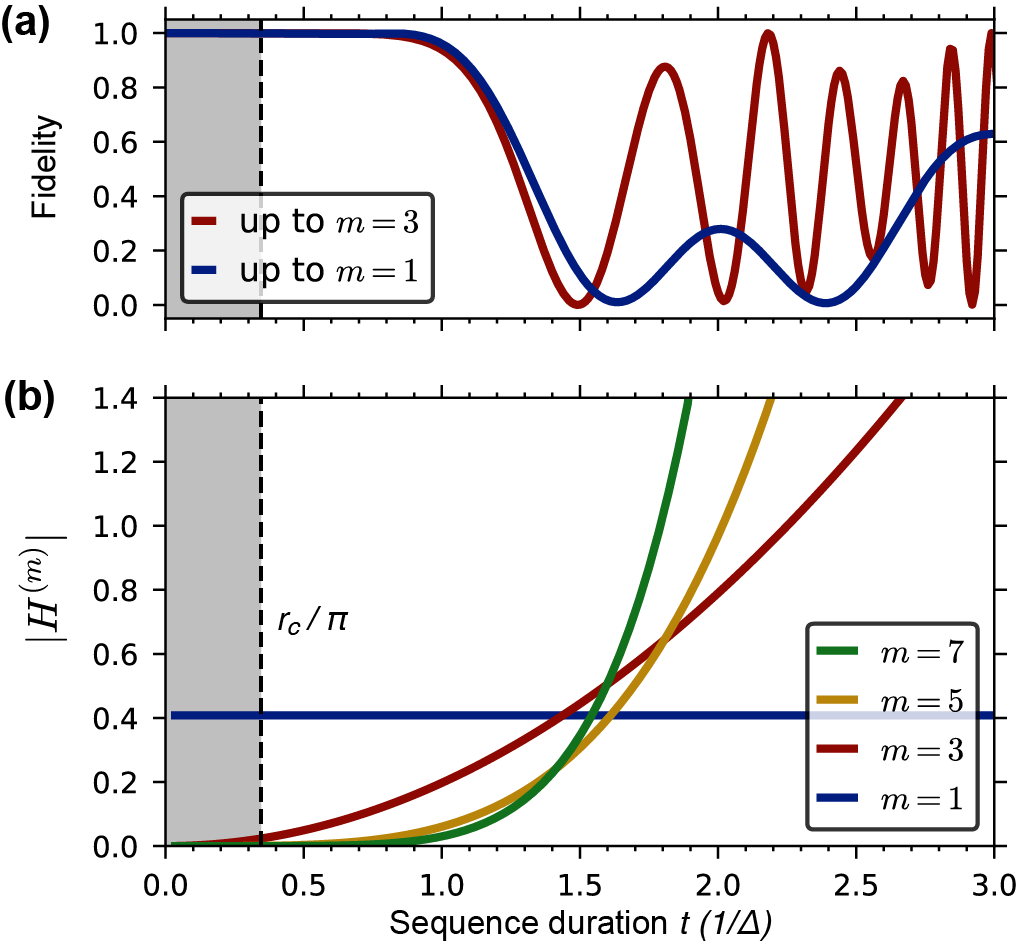}
    \caption{
 Fig 2
(a) State fidelity between exact and AHT evolution 
for varying sequence duration $t$, where $\mathcal{U}^{m}(t)$ is calculated for $m=1,3$ and compared to the ground truth unitary $U_0(t)$. (b) Spectral norm of Magnus expansion terms $|\bar{H}^{m}|$ for $m=1,3,5,7$ for the WAHUHA protocol, for varying sequence duration $t$. The region of guaranteed Magnus series convergence $t < r_c/\pi$ is shaded in gray.}
\label{fig:2}
\end{figure}

To investigate the discrepancy between the AHT prediction of time evolution and the exact dynamics, we include higher order contributions to calculate $\mathcal{U}^{m}$, for the same target Hamiltonian and initial state $\ket{+x}$ as above. Fig. \ref{fig:2}(a) shows a comparison of state fidelity between exact and AHT evolution \cite{Jozsa1994},
Notably, as the sequence length $t$ is extended, $\mathcal{U}^{(3)}$ produces fringes with increasing precession frequency, which suggests that the  energy scale of $\bar H^{(3)}$ is increasing with time $t$. This observation prompts us to consider the conditions where the Magnus expansion is expected to be valid. 


Typically, sufficient conditions for convergence of the Magnus expansion are established using a radius of convergence $r_c$~\cite{Blanes2009, Casas2007}. For the pulse sequences described here, this condition can be summarized as 
\begin{equation}
\label{eq:radius_sum}
\sum_i^n \big| H_i\big| \tau_i < \frac{r_c}{2 \pi},
\end{equation}
where $|\cdots|$ represents the spectral norm. 
The pulses generate a Hamiltonian that is piecewise constant during each free evolution interval, and the spectral norm  $|H_i| = \Delta/2$ remains unchanged. Using the upper bound $r_c \approx 1.087$ from Refs.~\cite{Blanes1998, Moan1999}, the condition in Eq.~(\ref{eq:radius_sum}) simplifies to
\begin{equation}
\label{eq:radius}
 \Delta \cdot t < \frac{r_c}{\pi} \approx 0.346.
\end{equation}
In order to guarantee convergence, the energy scale $\Delta$ of the interaction frame Hamiltonian and the total sequence duration $t$ can be chosen so their product does not exceed this threshold.

Although Eq.~(\ref{eq:radius}) assures convergence for the series expansion as a whole, it is common practice to truncate the series to the first few terms due to the increasing complexity of the higher order Magnus expansion.
To compare the relative contributions from these leading terms, 
the spectral norms $|\bar H^{(m)}|$ for the WAHUHA sequence are plotted in
Fig \ref{fig:2}(b) as a function of sequence duration $t$, for $m$ up to 7~\cite{Arnal2018, SI}.
As expected from Eq.~(\ref{eq:Hterms}), $|\bar H^{(1)}|$ is constant with respect to the sequence duration $t$, since the weighted average of the toggling frame Hamiltonians is unchanged.
However, the norms of higher order contributions increase with $t$, and eventually dominate.
Specifically, we observe that $|\bar H^{(m)}| \propto (\Delta \cdot t)^{m-1}$, which follows from the $m$ nested summations (or more generally, time integrals) that are present in $\bar H^{(m)}$. 

With these results in mind, we consider the typical operating regime for sensing, with NV centers in diamond as an example. These electronic spin defects are generally dominated by magnetic disorder within the diamond host, with spin resonance linewidths ranging from $\sim$$10^4$ Hz to upwards of $10^6$ Hz, which inform the typical magnitude of observable changes in $\Delta$. Moreover, the presence of NV hyperfine interactions with both onboard nitrogen and neighboring $^{13}$C nuclear spins in the diamond lattice often result in additional $10^6$ Hz-scale frequency detunings during control pulses, which may even be exploited to improve signal contrast~\cite{Arunkumar2023}. Optimal sequence durations $t$ for sensing are typically limited by dephasing/decoherence times that frequently exceed $10^{-6}$~s. In short, pulsed sensing experiments in these systems routinely operate in a regime where Magnus series convergence is not guaranteed.

\section{Exact Sensor Coupling Factor}
Considering the limitations of AHT outside the radius of convergence, we propose a method to calculate the exact (dimensionless) coupling factor, denoted here as $\alpha_\epsilon$. First, we consider deviations from the ground-truth unitary evolution by introducing a perturbation field $V$ (i.e., a target signal field to be measured) to the internal Hamiltonian $H_0$ in Eq.~(\ref{eq:U0}). The resulting unitary evolution operator is given by
\begin{equation}
\begin{aligned}
\label{eq:U1}
U_1 = \left(e^{- i (H_0 + V_n) \tau_n} P_n\right) \cdots \left(e^{- i (H_0 + V_1) \tau_1} P_1\right).
\end{aligned}
\end{equation}
We choose $V$ to be a small static (DC) perturbation field such that $V_k =\epsilon S_z$, with $\epsilon$ as a tunable scalar coefficient. Next, we define $U_\epsilon$ such that  $ U_1 = U_\epsilon U_0 $, or equivalently,
\begin{equation}\label{eq:Ueps}
U_\epsilon = U_1 U_0^\dagger.
\end{equation}
Note that $U_1 = U_0$ in the absence of a signal field ($\epsilon=0$), reducing $U_\epsilon$ to the identity operator. Therefore, any non-trivial rotation generated by $U_\epsilon$ is due to a nonzero signal contribution from $V$. 

Since $ U_\epsilon$ is unitary, it can be expressed in an exponential form $U_\epsilon = \exp^{- i \Omega(t)}$, where $\Omega(t)$ denotes an effective time-dependent operator to be determined. 
After diagonalizing $U_\epsilon$ to give the operator $\mathcal{D}(t) = \mathcal{P}^{-1}U_\epsilon  \mathcal{P}$ 
, we evaluate the principal logarithm of its eigenvalues along the diagonal of $\mathcal{D}(t)$ to obtain the exponent
\begin{equation}
\label{eq:generatorlog}
\Omega(t) = i \mathcal{P} \log[\mathcal{D}(t)]  \mathcal{P}^{-1}.
\end{equation}
Importantly, this expression for $\Omega(t)$ is exact, as long as the eigenvalues of $\Omega(t)$ are sufficiently small.
This is easily achieved by tuning the scale of the perturbation via $\epsilon$, for example, by ensuring $|\epsilon t | \ll 1$~\cite{Higham2008}. 
In line with AHT, we then extract an effective Hamiltonian $H_\epsilon$ by setting $\Omega(t) = H_\epsilon t$. However, we allow $H_\epsilon$ to retain a functional dependence on the sequence duration $t$. After normalizing the projection of $H_\epsilon(t)$ with respect to the perturbation, we obtain the exact coupling factor
\begin{equation}
\label{eq:modfactor_epsilon}
\alpha_\epsilon(t) = \frac{\sqrt{\sum_{i \in \{x,y,z\}} \text{Tr}(H_\epsilon^\dag S_i)^2}}{\left|\text{Tr}(V^\dag V)\right|}.
\end{equation}

As a check, we apply this method to a Ramsey sequence, which consists of a single toggling frame Hamiltonian $H_1 = S_z$. In this case, Eq.~(\ref{eq:U1}) simplifies to $U_\epsilon = e^{-i(S_z + \epsilon S_z)\tau}e^{+i(S_z )\tau} = e^{-i\epsilon S_z \tau}$ to reveal the effective Hamiltonian $H_\epsilon(t) = \epsilon S_z$, resulting in an exact coupling factor of $\alpha_\epsilon=1$ that agrees with AHT. 

For the WAHUHA sequence, we plot $\alpha_\epsilon(t)$ as a function of sequence duration $t$ in Fig 3, along with individual projections along $S_x, S_y, S_z$, for a nonzero initial detuning of 1 MHz. First, we note that $\alpha_\epsilon(t) = 1/\sqrt{3}$ as the sequence duration approaches zero, agreeing with the predictions due to leading order AHT in this regime. This result is expected, as we have shown in Fig. \ref{fig:2} that AHT yields the correct unitary propagator for sufficiently small timescales. However, as a function of sequence duration, $\alpha_\epsilon(t)$  exhibits oscillations between values of $1/3$ and $1/\sqrt{3}$.
Therefore, achieving the dimensionless coupling factor $1/\sqrt{3}$ (predicted by $\mathcal{H}^{(1)}$) requires careful choice of the sequence duration with respect to the energy scale of the interaction frame Hamiltonian $\Delta$.

\begin{figure}[t]
    \centering
\includegraphics[width=\linewidth]{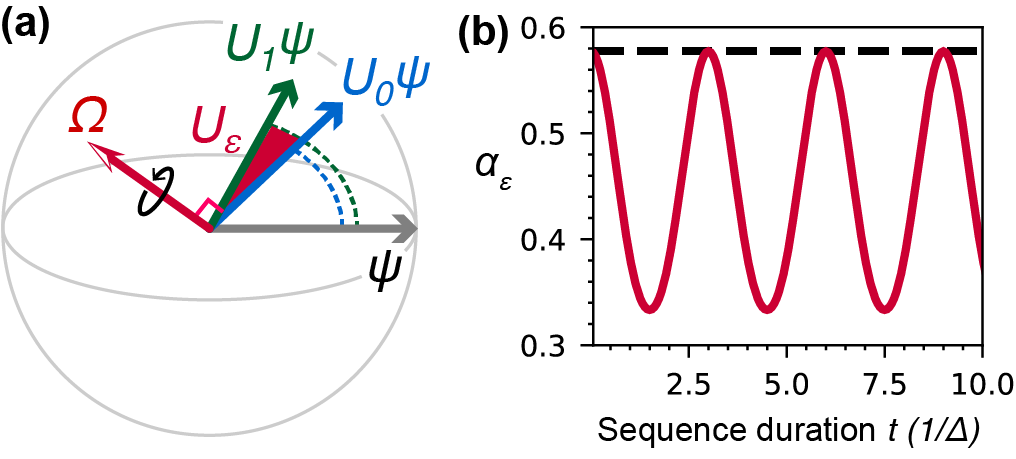}
    \caption{(a) An initial state $\psi$ is transformed by a pulse sequence unitary propagator to $U_0 \psi$. In the presence of a perturbation field $V$  (i.e., a target signal field to be measured), the final state is $U_1 \psi$. The operator $\Omega$ generates a unitary operator $U_\epsilon=e^{-i\Omega t}$ such that $U_1 = U_\epsilon U_0$. (b) Exact coupling factor $\alpha_\epsilon$ for the WAHUHA pulse sequence, as sequence duration $t$ is increased.
}
    \label{fig:3}
\end{figure}

\section{Retaining Convergence with Rapid Echoes}
\begin{figure}[t]
    \centering
\includegraphics[width=\linewidth]{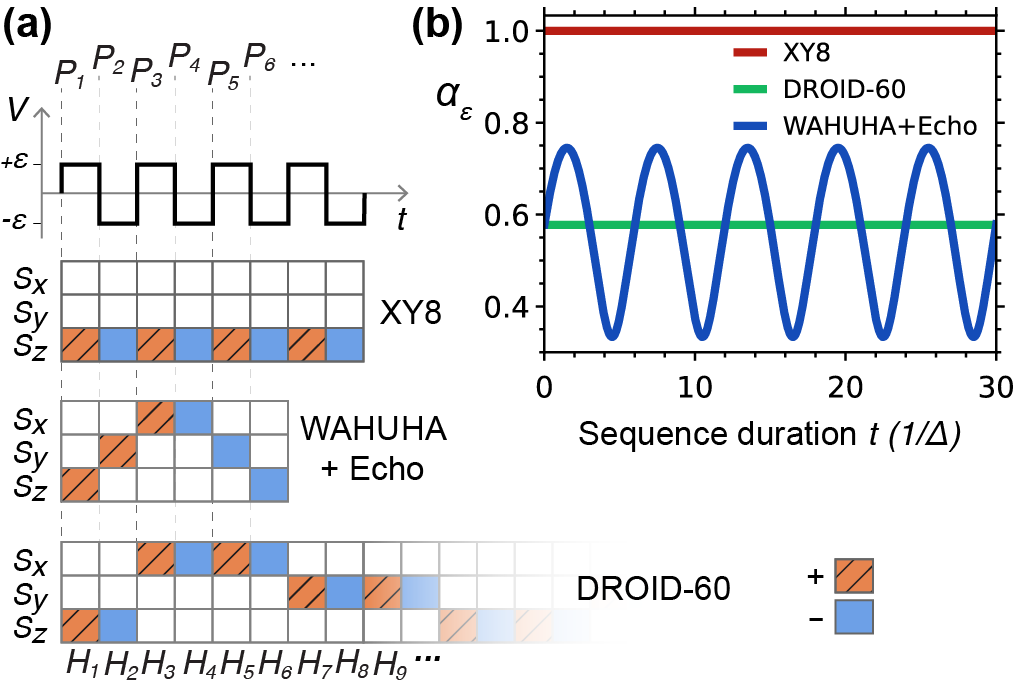}
    \caption{(a) Square wave perturbation with amplitude $\epsilon$, changing sign with each toggling frame. Each pulse $P_i$ is applied to the quantum sensor (e.g., NV electronic spins in diamond) before the $i$-th frame. For an initial sensor spin Hamiltonian $S_z$, the resulting toggling frame Hamiltonians for pulse sequences XY8, WAHUHA+Echo and DROID-60 are shown. (b) For each sequence, the exact coupling factor $\alpha_\epsilon$ is plotted as a function of sequence duration $t$. Calculations for XY8 and DROID-60 agree with predictions obtained from first-order AHT, i.e., $\bar H^{(1)}$.}
    \label{fig:4}
\end{figure}

Next, we extend this method to determine $\alpha_\epsilon$ in response to an oscillating (AC) target signal field. For simplicity, we describe the target AC field as a square wave perturbation that is present during the free evolution intervals, as illustrated in Fig. \ref{fig:4}(a).
By substituting $V_k = (-1)^k \epsilon S_z$ into Eq.~(\ref{eq:U1}) and following the steps in Eqs. (\ref{eq:Ueps}-\ref{eq:modfactor_epsilon}), we determine the effective time-dependent operator $\Omega$ due to this AC perturbation, and the corresponding $\alpha_\epsilon$. 

We then apply this analysis to three pulse sequences XY8, DROID-60 and WAHUHA+Echo \cite{Choi2020}. Their toggling frame transformations are shown in Fig \ref{fig:4}(a), for an initial static magnetic field along $+S_z$. All three sequences compensate each free precession interval around a given axis with an equal period in the opposite direction. Such ``echo" structures generate a null first order average Hamiltonian, $\bar H^{(1)} = 0$, suppressing contributions from the static field but retaining a response to the target AC magnetic field. 

In particular, XY8 and DROID-60 are commonly employed for high sensitivity AC magnetometry in NV ensembles \cite{Arunkumar2023}. 
In the case of XY8, which consists of a regular series of $\pi$ pulses, all toggling frame Hamiltonians commute ($[H_i, H_j] = 0$ for all $i, j$) to give $\alpha_\epsilon=1$, analogous to the Ramsey example discussed above.
In contrast, both DROID-60 and WAHUHA+Echo cycle between three orthogonal field orientations
to give the first-order AHT prediction $\alpha=1/\sqrt{3}$ in response to an AC target signal field~\cite{Zhou2020}. For each sequence, the exact coupling factor $\alpha_\epsilon$ is shown in \ref{fig:4}(b) as a function of sequence duration. Although all three sequences retain a response to AC magnetic fields, only AHT predictions for XY8 and DROID-60 align with the exact calculation ($\alpha_\epsilon(t) \approx \alpha$). Here,
exiting the radius of convergence established in Eq.~(\ref{eq:radius}) does not signal a breakdown of the AHT prediction.
However, in the case of WAHUHA+Echo, $\alpha_\epsilon$ clearly oscillates as $t$ is increased, deviating from the predictions of AHT.

We identify a common property of XY8 and DROID-60 that ensures the accuracy of AHT beyond the radius of convergence, namely the presence of a ``rapid" echo structure.
Specifically, we refer to rapid echoes as having pairs of nearest-neighbor toggling frame Hamiltonians with opposite sign, such that 
\begin{equation}
\label{eq:rapidechoes}
H_{2l} = -H_{2l-1}
\end{equation}
for all positive integers $l$. 
For any sequence that obeys this rapid echo structure, we show that all Magnus expansion contributions are zero (see Supplementary Material~\cite{SI}). The inclusion of a small AC signal field only introduces a $\sim$$(\epsilon t)^m$ contribution into each term $\bar H^{(m)}$; thus the system dynamics can be effectively described using only $\bar{H}^{(1)}$.
To summarize, Hamiltonian dynamics that commute at all times (such as with XY8 and Ramsey) and the presence of rapid spin echoes (DROID-60 and XY8) are examples of special cases where the Magnus expansion remains accurate beyond the sufficient radius of convergence.

\section{Outlook}
Recent applications of average Hamiltonian theory (AHT) range from robust dynamical decoupling and sensing protocols to engineering a wide array of Hamiltonians for probes of many-body physics~\cite{Martin2023, Lei2024arxiv}.
In this work, we characterize the breakdown of AHT for approximating features of time series data and the sensor coupling factor, illustrated by example pulse sequences used for sensing with solid-state spins. 
We also analyze sufficient convergence criteria for the Magnus expansion for a general two-level system.

Using exact methods to determine the sensor coupling factor, we show that rapid echoes allow the Magnus approximation to remain accurate well beyond its guaranteed radius of convergence. 
This beyond AHT result is merely one example of structures that validate the Magnus approximation, opening up opportunities for Hamiltonian engineering principles without the increasing computational cost of higher order Magnus expansion terms \cite{Tyler2023, Zhou2024}.

These exact methods are valid regardless of Magnus expansion convergence, and can be readily adapted to more complex systems and sources of noise. 
Example future areas of application may include finite pulse errors, rotation angle errors, and interacting systems, which can be major contributors to dephasing and decoherence. 
In addition, frequency domain analysis can complement existing filter function approaches, which are frequently employed to study dynamical decoupling protocols across quantum information science and related fields~\cite{Cywinski2008, Uhrig2008, Biercuk2009}.
We foresee future use of beyond AHT techniques in designing optimized measurement protocols that adhere to leading order Magnus approximations in the presence of dephasing sources, and as potential reward metrics for optimal control and machine learning approaches~\cite{ Poggiali2018, OKeeffe2019, Peng2022}.
Extending this beyond AHT approach to higher-dimensional systems may also be relevant to sensing protocols that leverage qudit structures~\cite{Mamin2014, Hart2021, Zhou2023}. However, in such cases we expect the contributions to the radius of convergence in Eq.~(\ref{eq:radius}) to remain pertinent.

\begin{acknowledgments}
We gratefully acknowledge Hengyun Zhou and Haoyang Gao for helpful discussions. This work is supported by, or in part by, the DEVCOM Army Research Laboratory under Contract No. W911NF1920181 and under Cooperative Agreement Number W911NF-24-2-0044 (SCC); the DEVCOM ARL Army Research Office under Grant No. W911NF2120110; the U.S. Air Force Office of Scientific Research under Grant No. FA9550-22-1-0312; and the University of Maryland Quantum Technology Center.
\end{acknowledgments}


\bibliographystyle{apsrev4-1}
\bibliography{AHT.bib}
\end{document}


\title{Beyond Average Hamiltonian Theory for Quantum Sensing}
\date{\today}

\author{Jner Tzern Oon}
\affiliation{\UMDphy}
\affiliation{\QTC}

\author{Sebastian C. Carrasco}
\affiliation{\ARL}

\author{Connor A. Hart}
\affiliation{\QTC}

\author{George Witt}
\affiliation{\UMDphy}

\author{Vladimir S. Malinovsky}
\affiliation{\ARL}

\author{Ronald Walsworth}
\affiliation{\UMDphy}
\affiliation{\QTC}
\affiliation{\UMDEECS}

\date[]{}

\maketitle

\tableofcontents

 
\section{Magnus Expansion}

The Magnus expansion is an approximate solution to a linear differential equation of the form
\begin{equation}
\psi'(t) = O(t) \psi(t)\,,
\end{equation}
where the solution is given by the expression
\begin{equation}
\psi(t) = \exp{\left(\Omega(t)\right)}\,.
\end{equation}
For the pulse sequences considered in the main text, $O(t) = -i H_k$ during the $k$-th period of free evolution ($(k-1)\tau \leq t < k \tau$, with $\tau$ being the duration of each period). $H_k$ denotes the toggling frame Hamiltonian during this interval. 
To evaluate the  $m$-th order term in the Magnus series,  $\Omega_m$, we follow the steps outlined in Arnal et al. 2018 \cite{Arnal2018},
\begin{equation}
    \Omega_m = \frac{1}{m} \sum_\sigma (-1)^{d_b} \frac{d_a! d_b!}{m!} A(\sigma) = \sum_\sigma g(\sigma) A(\sigma)\,,
\end{equation}
where
\begin{equation}
    \label{eq:Asigma}
    A(\sigma) = \int_0^t dt_1 \int_0^{t_1} dt_2 \cdots \int_0^{t_{m-1}} dt_m O(t_{\sigma(1)}) O(t_{\sigma(2)}) \cdots O(t_{\sigma(m)})\,.
\end{equation}
Here, $\sigma$ is a permutation of $\{1, \ldots, m\}$, and $d_a$ and $d_b$ is the number of ascents and descents in a permutation $\sigma$.

\section{Discrete Time}
We consider a series of toggling frame Hamiltonian operators which are constant during periods of duration $\tau$. For a sequence of total duration $n\tau$, Eq. \eqref{eq:Asigma} can be expressed by discrete sums over these $n$ sub-intervals, 
\begin{equation}
\begin{aligned}
A(\sigma) &= 
\sum_{x_1=1}^{n}\sum_{x_2=1}^{x_1}\dots \sum_{x_{m}=1}^{x_{m-1}} 
\int_{(x_1-1)\tau}^{x_1\tau} dt_1
\int_{(x_2-1)\tau}^{\text{min}[t_2, x_2\tau]} dt_2 \cdots
\int_{(x_m-1)\tau}^{\text{min}[t_{m-1}, x_m\tau]} dt_m
O(t_{\sigma(1)}) \cdots O(t_{\sigma(m)})\, .
\end{aligned}
\end{equation}
Setting $A(\sigma) = \sum_{\vec{x}} a(\vec{x}, \sigma)$ with
\begin{align}
a(\vec{x},\sigma) &= \int_{(x_1-1)\tau}^{x_1\tau} dt_1\int_{(x_2-1)\tau}^{\text{min}[t_2, x_2\tau]} dt_2 \cdots
\int_{(x_m-1)\tau}^{\text{min}[t_{m-1}, x_m\tau]} dt_m
O(t_{\sigma(1)}) \cdots O(t_{\sigma(m)}) \\
&= O_{x_{\sigma(1)}} \cdots O_{x_{\sigma(m)}} \int_{(x_1-1)\tau}^{x_1\tau} dt_1\int_{(x_2-1)\tau}^{\text{min}[t_2, x_2\tau]} dt_2 \cdots
\int_{(x_m-1)\tau}^{\text{min}[t_{m-1}, x_m\tau]} dt_m \\
&= \mathcal{O}_{\vec x, \sigma} f(\vec x) \, ,
\end{align}
where $O_{k} = O((k-1)\tau\leq t<k\tau)$, the function $f(\vec x)$ is the result of the nested integrals, and $O_{x_{\sigma(1)}} \cdots O_{x_{\sigma(m)}} = \mathcal{O}_{\vec x, \sigma}$.
We can evaluate the individual contributions for an arbitrary sub-interval denoted by $\vec{x} = (x_1, \dots, x_m$). During this sub-interval, $O(t_{\sigma(1)}) \cdots O(t_{\sigma(m)}) = \mathcal{O}_{\vec x, \sigma}$ is constant and can be taken out of the integral. 
Since $x_i \geq x_{i+1}$, the sequence $\vec{x}$ is in descending order.
Therefore, for a pair of neighboring indices $x_i$ and $x_{i+1}$, the
corresponding nested integrals in $a(\vec{x},\sigma)$ fall under one of two cases:
\begin{equation}
\int_{(x_i-1)\tau}^{\text{min}[t_{i-1}, x_i\tau]}  dt_i \int_{(x_{i+1}-1)\tau}^{\text{min}[t_{i}, x_{i+1}\tau]} dt_{i+1} = 
\ \begin{dcases}
\int_{(x_i-1)\tau}^{\text{min}[t_{i-1}, x_i\tau]}  dt_i \int_{(x_{i+1}-1)\tau}^{ x_{i+1}\tau} dt_{i+1},
&\text{if }x_i > x_{i+1}\\
\int_{(x_i-1)\tau}^{\text{min}[t_{i-1}, x_i\tau]}  dt_i \int_{(x_{i+1}-1)\tau}^{ t_i} dt_{i+1},
&\text{if }x_i = x_{i+1}
\end{dcases} \, .
\end{equation}
When $x_i > x_{i+1}$, the bounds of the integrals are independent, and can be evaluated separately. However, if $x_i = x_{i+1}$, we combine and evaluate the integrals that correspond to a subsequence containing only repeated indices. For $k$ repeated indices, i.e., $x_{i+1}= x_{i+2} = \dots = x_{i + k}$, we obtain the following result:
\begin{equation}
\int_{(x_{i+1}-1)\tau}^{ x_{i+1}\tau}  dt_{i+1} 
\int_{(x_{i+2}-1)\tau}^{ t_{i+1}} dt_{i+2} \cdots
\int_{(x_{i+k}-1)\tau}^{ t_{i+k-1}} dt_{i+k} 
= \frac{ \tau^k}{k!} \, .
\end{equation}
When applied to the entire set of indices $\vec{x}$, each unique integer $x_j$ with $k_j$ repeated entries introduces a multiplicative factor $\tau^{k_j}/(k_j!)$ during evaluation of the corresponding time integrals. For example, for a particular set of $m=11$ indices $\vec{x}' = (4,4,3,3,3,3,3,2,1,1,1)$, we obtain
\begin{equation}
a(\vec{x}',\sigma) = 
\underbrace{\frac{\tau^{11}}{2!\,5!\,1!\,3!}
}_{f\left(\vec{x}'\right)}
\quad
\underbrace{O_{x'_{\sigma(1)}}  O_{x'_{\sigma(2)}}\cdots  O_{x'_{\sigma(11)}}}_{\mathcal{O}_{\vec{x}',\sigma}} \, ,
\end{equation}
%
where $\tau_i$ describes any time between $i\tau \leq t<(i+1)\tau$ when $O(t)$ is constant.
As a result, the Magnus expansion can be expressed as
\begin{equation}
\begin{aligned}
\Omega_m  &= \sum_{\sigma,\vec{x}} g(\sigma) a(\vec{x}, \sigma)\\
&= \sum_{\sigma,\vec{x}} g(\sigma) f(\vec{x}) \mathcal{O}_{\vec{x}, \sigma} \, .
\end{aligned}
\end{equation}
%



\section{Magnus expansion is zero for rapid echoes}

We have that 
\begin{equation}
    a(\vec{x},\sigma) = \sum_{\vec{x}} f(\vec{x}) \mathcal{O}_{\vec{x}, \sigma} \, ,
\end{equation}
where $\vec x=(x_1, x_2, \ldots, x_m)$ is a vector of indices with $x_j \geq x_{j+1}$. 

We define rapid echos as having each odd pulse followed by an even pulse that generates a toggling frame Hamiltonian of opposite sign, such that $O_{2j-1}=-O_{2j}$.
To show that $a(\vec{x},\sigma)$ is zero for an arbitrary sequence of rapid echoes, we first consider the case where $x_1$ is an even number, repeats $k$ times ($x_1 = x_2 =\dots=x_k$), and is then followed by the index $x_{k+1}$ that doesn't belong to the same echo ($x_{k+1} \neq x_1$ and $x_{k+1} \neq x_1-1$), namely
\begin{equation}
    \vec{x}_0=(\underbrace{x_1, \ldots, x_1}_{\text{$k$ repetitions}}, x_{k+1}, \ldots) \, .
\end{equation}
We shall see that $a(\vec{x}_0,\sigma)$ is cancelled by sets of indices of the form
\begin{equation}
    \vec{x}_l=(\underbrace{x_1, \ldots, x_1}_{\text{$k-l$ repetitions}}, \underbrace{x_1-1, \ldots, x_1-1}_{\text{$l$ repetitions}}, x_{k+1}, \ldots) \, .
\end{equation}
Specifically, $a(\vec{x}_0,\sigma)$ is cancelled by all the other sets of indices that start with $k$ pulses belonging to the same echo. To show this, we notice that
\begin{equation}
    f(\vec{x}_l) = \frac{1}{l! (k-l)!} f(x_{k+1}, \ldots) = \frac{1}{k!} \binom{k}{l} f(x_{k+1}, \ldots) \, ,
\end{equation}
which results in
\begin{equation}
    \sum_{l=0}^k f(\vec{x}_l) \mathcal{O}_{\vec{x}_l, \sigma} = \frac{1}{k!} f(x_{k+1}, \ldots) \sum_{l=0}^k (-1)^l \binom{k}{l} \mathcal{O}_{\vec{x}_0, \sigma} \, .
\end{equation}
The factor $(-1)^l$ comes from exchanging each $O_{x_1-1}$ for $-O_{x_1}$
as $\mathcal{O}_{\vec{x}_l, \sigma}$ is converted to $\mathcal{O}_{\vec{x}_0, \sigma}$. Applying the identity
\begin{equation}
    \sum_{j=0}^i (-1)^j \binom{i}{j} = 0\, ,
\end{equation}
we obtain
\begin{equation}
\sum_{l=0}^k f(\vec{x}_l) \mathcal{O}_{\vec{x}_l, \sigma}= 0 \, .
\end{equation}
This implies that the total contribution from all set of indices $\vec{x}$ starting with $k$ pulses in the same echo is equal to zero. Therefore, since the sum $\sum_{\vec{x}} f(\vec{x}) \mathcal{O}_{\vec{x}, \sigma}$ can be sub-divided according to the number of starting pulses belonging to the same echo, we immediately see that
\begin{equation}
    \sum_{\vec{x}} f(\vec{x}) \mathcal{O}_{\vec{x}, \sigma}  = \sum_{\substack{\vec{x}\\ k=1}} f(\vec{x}) \mathcal{O}_{\vec{x}, \sigma}  + \sum_{\substack{\vec{x}\\ k=2}} f(\vec{x}) \mathcal{O}_{\vec{x}, \sigma} + \cdots = 0 \, .
\end{equation}
Under these conditions, we conclude that the Magnus series contribution at arbitrary order is zero, i.e.,
\begin{equation}
\Omega_m  
= \sum_{\sigma,\vec{x}} g(\sigma) a(\vec{x}, \sigma) 
= \sum_{\sigma} g(\sigma) \sum_{\vec{x}}f(\vec{x}) \mathcal{O}_{\vec{x}, \sigma} 
= 0 \, .
\end{equation}

\section{Numerical simulation with random echo sequences}

As a sanity check, we generated $10^4$ random sequences of $n=20$ with rapid echo structure, computed the Magnus expansion up to order $m=5$, and calculated the spectral norm of each order. We obtained that all the spectral norms are around $10^{-16}$, around machine precision (for double-precision numbers) and therefore zero for practical considerations. We conclude that no rapid echo sequences have non-zero Magnus expansion for these parameters, consistent with the result in the previous section. 

The random echo sequences are described by the most general Hamiltonian (up to a multiplicative constant) for a two-level system, $H(\theta, \phi) = \sin \theta \cos \phi \sigma_x + \sin \theta \sin \phi \sigma_y + \cos \theta \sigma_z$. To generate a random Hamiltonian, we select a random value of $\theta$ within $[0, \pi]$ and a random value of $\phi$  within $[0,2\pi]$. A sequence of $M$ rapid echoes ($n=2M$) is generated using $M$ random Hamiltonians, each repeated with the opposite sign. Consequently, the sequence is $H(\theta_1, \phi_1)$, $-H(\theta_1, \phi_1)$, \ldots, $H(\theta_M, \phi_M)$, and $-H(\theta_M, \phi_M)$.

\bibliographystyle{apsrev4-1}
\bibliography{AHT.bib}